\documentclass[pre,onecolumn,nofootinbib,superscriptaddress,showkeys,11pt]{revtex4}
\usepackage{comment}
\usepackage{graphicx}
\usepackage{graphicx,subfigure}
\usepackage{amsmath,amssymb,amsfonts}

\usepackage[usenames]{color}
\usepackage{bm}
\usepackage{epsfig}
\usepackage{dcolumn}
\usepackage{xspace}
\usepackage{orcidlink}

\begin{document}
\title{A generalized framework for viscous and non-viscous damping models}
\author{Soumya Kanti Ganguly\,\orcidlink{0009-0000-0852-7876}}
\affiliation{Ongil Private Limited, Chennai 600113, India}
\author{Indrajit Mukherjee\,\orcidlink{0000-0001-6840-929X}\protect\footnote
{E-mail address: imukherjee@aero.iiests.ac.in}}
\affiliation{Department of Aerospace Engineering and Applied Mechanics, IIEST, Shibpur, Howrah 711103, India}
\begin{abstract}
The inadequacy of the classical viscous damping model in capturing dissipation across a wide range of applications has led to the development of 
non-viscous damping models. While non-viscous models describe damping force satisfactorily, they offer limited physical insight. Leveraging an 
existing framework, well known to the physics community, this article provides fresh insights into the framework of non viscous damping for engineers. 
For this purpose, we revisit the motion of a particle coupled to a bath of harmonic oscillators at a finite temperature. We obtain a general expression 
for non-viscous damping in terms of a general memory kernel function. For specific choices of the kernel function, we derive exact expressions for a host 
of non-viscous damping models including the classical viscous damping as a special case.
\end{abstract}
\maketitle
\section{Introduction}
Recent advances in different fields of engineering ranging from  aerospace, civil, mechanical to naval architecture have prompted the development of 
various phenomenological models of damping as the classical viscous damping turns out to be inadequate in describing the mechanism of dissipation in 
diverse application areas arising in these fields \cite{Tsai}-\cite{Nashif}. The main feature of these linear, non-viscous damping models is that the 
instantaneous damping force depends on the past history of velocities through a convolution integral over a kernel function, as expressed in 
Eq.\eqref{nonviscous}

\begin{equation}\label{nonviscous}
F_{d}(t) = \int_{0}^{t}G(t-\tau)\dot{x}(\tau)d\tau,
\end{equation}

where $F_{d}$ is the instantaneous  damping force, $G(t)$ is the damping kernel function and $\dot{x}(t)$ is the velocity of the system. Non-viscous 
damping is also commonly referred to as viscoelastic damping. A succinct account of these models can be found in Adhikari \cite{Adhikari0}, 
and the references therein. Some of the popular non-viscous damping models along with the Fourier Transform of their damping kernel are enlisted in 
TABLE.\ref{TABLE}. While these phenomenological models satisfactorily describe the dissipative forces, they are not developed from a first principle 
based approach. These damping models are approximate mathematical representations of dissipation and do not provide a detailed explanation of the 
underlying physics of the dissipation mechanism. The present article addresses this gap by taking recourse to a physics based approach, grounded in 
the principle of statistical mechanics, to expound the phenomenological models of non-viscous damping models. For this purpose, we revisit the problem 
of damping by adopting a classic approach by Zwanzig \cite{ZwanzigA}. Although his formulation was well-suited for pedagogical purposes, it effectively 
conveys the core aspects of the problem. In the following sections, we will provide a brief overview of Zwanzig's work, and derive a general expression 
for the memory kernel related to the viscous term. Subsequently, we will present the well-known forms of the kernel for various dissipation models, 
followed by some concluding remarks.

\begin{table}[h]\label{TABLE}
\centering
\begin{tabular}{cccc}
\hline
 & \textbf{Model} & $G(\omega)$ & \textbf{References} \\
\hline
 & Biot model & $\sqrt{\frac{2}{\pi}}\sum_{k=1}^{n}\frac{a_{k}b_{k}}{b^{2}_{k} + \omega^{2}}$ & Biot \cite{Biot}\\
 & Caputo model & $(\pi\omega)^{-1/2}$ & Caputo \cite{Caputo},\cite{Bagley2}\\
 & BT model & $\frac{G_{0} + G_{\infty}\omega^{\alpha}}{1 + \omega^{\beta}}$ & Bagley \& Torvik \cite{Bagley1}\\
 & GHM model &  
 $\sqrt{\frac{2}{\pi}}\sum_{k=1}^{n} a_{k}\Big{\{} \frac{1}{b^{2}_{1k} + \omega_{k}^{2}} - \frac{1}{b^{2}_{2k} + \omega_{k}^{2}} \Big{\}}\Big]$  
 & Golla, McTavish \& Hughes \cite{Hughes1,Hughes2}\\
 & Gaussian model & $\exp(-\omega^{2}/4\mu)/\sqrt{2\mu} $ & Adhikari \& Woodhouse \cite{Adhikari1} \\
 & Chatterjee model & $(2/\pi)^{1/2}\Gamma(3/4)\sin(\pi/8)\vert\omega\vert^{-3/4}$ & Chatterjee \cite{Chatterjee}\\
\hline
\end{tabular}
\caption{Non-viscous damping models with memory kernel}
\label{tab:example}
\end{table}

\section{The environment as a bath of harmonic oscillators and the memory kernel}

Understanding the phenomenon of damping in physical systems has long been a priority for physicists. Schwinger \cite{Schwinger} was 
the first to study the motion of a quantum particle subjected to random forces using his action principle and Green's function techniques. 
Following this, Feynman and Vernon \cite{Feynman} used the Feynman path integral method to examine the motion of a quantum particle in a 
linear dissipative environment, where it was modeled by a bath of harmonic oscillators. The field continued to evolve through various 
contributions \cite{Ford}-\cite{Gardiner}, with R. Zwanzig \cite{ZwanzigA,ZwanzigB} offering a particularly simple classical Langevin framework 
for this problem. In this section we will outline Zwanzig's technique and derive a general expression for the memory kernel.

Let ($x,p$) be the position and momentum co-ordinates of a partcle of mass $m$. If its potential energy be $U(x)$, then the Hamiltonian
of the particle is given by

\begin{equation}\label{Zwanzig0}
H_{0}(x,p) = \frac{p^{2}}{2m} + U(x).   
\end{equation}

Let us now consider a bath of $N$ oscillators with unit mass, having position and momenta co-ordinates given by the set $\{X,P\}$.
The Hamiltonian for the oscillator bath interacting with the particle is given by

\begin{equation}\label{Zwanzig1}
H_{b}(X,P,x) = 
\sum_{j=1}^{N}\bigg[\frac{P_{j}^{2}}{2} + \frac{\omega^{2}_{j}}{2}\bigg(X_{j} - \frac{\gamma_{j}}{\omega^{2}_{j}}x\bigg)^{2} \bigg].   
\end{equation}

Note that the interaction between the particle and the bath of oscillators is captured by the second term in the R.H.S. of the
above equation. For each oscillator, the minima of the potential energy is shifted by $\gamma_{j}x/\omega^{2}_{j}$. The total 
Hamiltonian $H = H_{0} + H_{b}$, which gives us the following equations of motion for the particle and the bath degrees of freedom.

\begin{eqnarray}\label{Zwanzig3}
\frac{dx}{dt} &=& \frac{p}{m}, \quad  \frac{dp}{dt} = -\frac{dU(x)}{dx} + 
\sum_{j=1}^{N}\gamma_{j}\bigg(X_{j} - \frac{\gamma_{j}}{\omega^{2}_{j}}x\bigg), \\ \nonumber  
\textrm{and} \quad 
\frac{dX_{j}}{dt} &=& P_{j}, \quad  \frac{dP_{j}}{dt} = -\omega^{2}_{j}X_{j} + \gamma_{j}x
\end{eqnarray}

The solutions for the oscillators are given by

\begin{equation}\label{Zwanzig4}
X_{j}(t) - \frac{\gamma_{j}}{\omega^{2}_{j}}x(t) = \bigg(X_{j}(0) - \frac{\gamma_{j}}{\omega^{2}_{j}}x(0)\bigg)\cos\omega_{j}t 
+ P_{j}(0)\frac{\sin\omega_{j}t}{\omega_{j}} - \gamma_{j}\int_{0}^{t}dt'\frac{p(t')}{m} \frac{\cos\omega_{j}(t-t')}{\omega_{j}^{2}}             
\end{equation}

Substituting this in the second equation in Eq.\eqref{Zwanzig3}, we get the following equation of motion for the particle:

\begin{equation}\label{Zwanzig5A}
\frac{dp(t)}{dt} = -\frac{dU(x)}{dx} - \int_{0}^{t}dt'G(t')\frac{p(t-t')}{m} + F(t).  
\end{equation}

\begin{equation}\label{Zwanzig5B}
\textrm{Where} \quad 
G(t) = \sum_{j=1}^{N} \frac{\gamma_{j}^{2}}{\omega^{2}_{j}}\cos\omega_{j}t,  \\ \nonumber
\end{equation}

is the memory kernel function, associated with the damping or velocity dependent friction term proportional to $p(t)/m$. And, the 
other quantity

\begin{equation}\label{Zwanzig5C}
F(t) = \sum_{j=1}^{N} \bigg[\gamma_{j}P_{j}(0)\frac{\sin\omega_{j}t}{\omega_{j}} - 
\gamma_{j}\bigg(X_{j}(0) - \frac{\gamma_{j}}{\omega^{2}_{j}}x(0)\bigg)\cos\omega_{j}t \bigg],
\end{equation}

is an external force due to the bath of harmonic oscillators. For the set of initial values $\{X_{j}(0),P_{j}(0)\}$, $F(t)$ can be 
determined at every $t$. However, this is practically impossible if $N$ is comparable to Avogadro's number. For such large $N$, it 
is justified to resort to a statistical description. Furthermore, the Central limit theorem allows us to sample the $\{X_{j}(0),P_{j}(0)\}$ 
values from identical and independent Gaussian distributions with means $\{\gamma_{j}\omega^{-2}_{j}x,0\}$, and a specified finite variance. 
If the oscillators are sampled from the Boltzmann distribution at some equilibrium temperature $T$, then

\begin{eqnarray}\label{Zwanzig6}
&&f_{b}(X,P) = \frac{\exp(-H_{b}/k_{B}T)}{Z_{B}},  \\ \nonumber
\textrm{where} \quad 
&&\int_{-\infty}^{+\infty} \int_{-\infty}^{+\infty} dX dP f_{b}(X,P) = 1.
\end{eqnarray}

The exponent $H_{b}$ is given by Eq.\eqref{Zwanzig1}, and $k_{B}$ is the Boltzmann constant, and $Z_{b}$ is the normalization constant
(see Eq.\eqref{App4}). The thermal average w.r.t. $f_{b}(X,P)$, by definition is given by:

\begin{equation}\label{App6}
\langle A \rangle = \prod_{i} \int_{-\infty}^{+\infty} \int_{-\infty}^{+\infty} dX_{i} dP_{i} f_{b}(X,P) A(X,P).
\end{equation}

Using this definition and Eq.\eqref{Zwanzig6}, we can calculate the following statistical averages:

\begin{eqnarray}\label{Zwanzig7}
\Big\langle \Big(X_{j}(0) - \gamma_{j}\omega^{-2}_{j}x(0)\Big) \Big\rangle &=& 0, \quad \langle P_{j}(0) \rangle = 0 \\ \nonumber
\Big\langle \Big(X_{j}(0) - \gamma_{j}\omega^{-2}_{j}x(0)\Big)^{2} \Big\rangle &=& \frac{k_{B}T}{\omega^{2}_{j}}, 
\quad \langle P^{2}_{j}(0) \rangle = k_{B}T, \quad \forall j = 1(1)N. 
\end{eqnarray}

Where $\langle A(X,P) \rangle$ denotes the ensemble average of the canonical variable $A$ w.r.t. $f_{b}(X,P)$. Similarly,
the autocorrelation for $F(t)$ is given by (see Eq.\eqref{App7}-\eqref{App10}):

\begin{equation}\label{Zwanzig8}
\langle F(t)F(t') \rangle = k_{B}T G(t-t'). 
\end{equation}

Where, $G(t)$ is the same memory kernel function in Eq.\eqref{Zwanzig5B}  (see Eq.\eqref{App9}). The Eq.\eqref{Zwanzig8} is a 
fluctuation dissipation theorem, which suggests that the environment acts as a thermal reservoir, with fluctuating forces that 
serve as agents of dissipation on the moving particle, influenced by a temperature-dependent term. Note that the temperature 
dependence is due to the thermal nature of the reservoir, which is sampled according to the Boltzmann distribution. In the 
absence of the thermal effects, we could have chosen a different probability distribution. The history of the environment, 
extending to the present moment, is encapsulated in this dissipative term through a memory function $G(t)$. Each oscillator 
applies a force on the particle by coupling itself through $\gamma$. Owing to the size of the bath degrees of freedom, we 
ignore any kind of feedback due to the particle on the bath. In the thermodynamic limit, when the environment is infinitely 
large, the frequency of the oscillators become a continuous variable $\omega$, and may be assumed to be distributed according 
to some function $w(\omega)$. Then the kernel function $G(t)$ in Eq.\eqref{Zwanzig5B} is given by (see Eq.\eqref{App10})

\begin{equation}\label{Zwanzig9}
\textrm{or,} \quad 
G(t) = \int_{-\infty}^{\infty}d\omega w(\omega)\frac{\gamma^{2}(\omega)}{2\omega^{2}} e^{i\omega t}.             
\end{equation}

Note, that the harmonic oscillator formalism provides a very natural framework for expressing the different quantities in terms of their 
Fourier transforms. This is contrary to the usual practice within the engineering community, where Laplace transform is the natural choice. 
For all practical purposes, both choices are mathematically equivalent. If the environment is a solid medium, then the upper limit of the 
integral in Eq.\eqref{Zwanzig9} will be replaced by the Debye frequency. The Debye frequency is determined by the speed of sound in the solid 
medium. For the purpose of modeling an environment, and therefore the memory function, the physics of the problem is of central importance. 
This governs our choice of $\gamma^{2}(\omega)$, which must be analytic in $\omega$. A classic example of an environment having profound 
consequences is in the field of atomic physics. When an atom spontaneously decays from its excited state to its ground state by emitting 
electromagnetic radiation, this radiation can be significantly inhibited by suitably designing a cavity. Since cavities can sustain radiation, 
different cavities will have different densities of states, which can be regulated to control the lifetime of the atom in the excited state 
\cite{Kleppner}. In the next section, we will see how the different models for non-viscous damping tabulated in TABLE.\ref{TABLE} can be recovered 
by appropriately choosing $\gamma^{2}$. 

\section{Results and Discussion}

\subsection{Case 1: Classical viscous damping model}

If $w(\omega) = 2\omega^{2}$ and $\gamma^{2}(\omega) = \eta$, a constant also known as the coefficient of friction, then using the 
identity $\int_{-\infty}^{\infty}d\omega e^{i\omega t} = \delta(t)$, we get $G(t) = \eta\delta(t)$. If we put this in Eq.\eqref{Zwanzig5B}, 
we find

\begin{equation}\label{Zwanzig10A}
\int_{-\infty}^{t} dt'\eta\delta(t')\frac{p(t-t')}{m} = \eta\frac{p(t)}{m},                 
\end{equation}

which is the well known classical viscous damping term. The autocorrelation function for the environmental force $F(t)$ is given by   

\begin{equation}\label{Zwanzig10B}
\langle F(t)F(t') \rangle = \eta k_{B}T \delta(t-t'). 
\end{equation}

This suggests that the forces at different times are uncorrelated. In other words, the environment retains no memory of the 
system. Such an environment is known as a memoryless or a Markovian environment in the physics literature \cite{ZwanzigB,Gardiner}. 
For the other damping models, the environment will have a much more complicated non-Markovian behaviour.  
\subsection{Case 2: The Biot model (M.A. Biot)}

The Biot model has a memory kernel function which is exponentially decaying in time, but has many time-constants $\{a_{k}\}$. 
If we choose 

\begin{equation}\label{Zwanzig11A}
w(\omega) = 2\omega^{2}, 
\quad \textrm{and} \quad 
\gamma^{2}(\omega) = \sqrt{\frac{2}{\pi}}\sum_{k} \frac{a_{k}}{a^{2}_{k} + \omega^{2}}  
\end{equation}

The memory kernel function in Eq.\eqref{Zwanzig9} will then give us 

\begin{equation}\label{Zwanzig11B}
G(t) = \sum_{k=1}^{n} \exp(-a_{k}t)   
\end{equation}

\subsection{Case 3: Fractional derivative model (Caputo and Chatterjee)}

To derive the memory kernel for the fractional derivative model, let us choose \cite{Gradshteyn} 

\begin{equation}\label{Zwanzig7A}
w(\omega) = 2\omega^{2}, 
\quad \textrm{and} \quad 
\gamma^{2}(\omega) = \frac{(2/\pi)^{1/2}\sin(\pi z/2)}{\vert\omega\vert^{1-z}}, \quad \textrm{where} \quad  0 < \Re(z) < 1.	 	
\end{equation}

The memory kernel function in Eq.\eqref{Zwanzig9} will then give us 

\begin{equation}\label{Zwanzig7B}
G(t) = (2/\pi)^{1/2}\sin(\pi z/2)\int_{-\infty}^{\infty}d\omega\frac{e^{i\omega t}}{\vert\omega\vert^{1-z}}. 
\quad \textrm{Or,} \quad 
G(t) = \frac{t^{-z}}{\Gamma(1-z)} \quad  (0 < \Re(z) < 1).	 	
\end{equation}

Where $\Gamma$ is the Euler gamma function. If $z = 1/2$, then we obtain the well-known Caputo fractional derivative case

\begin{equation}\label{Zwanzig7C}
G(t) = \frac{1}{\Gamma(1/2)}t^{-1/2}.	 	
\end{equation}

Another intriguing case of $z = 1/4$, due to Chatterjee \cite{Chatterjee} can be derived by assuming \cite{Gradshteyn} 

\begin{eqnarray}\label{Zwanzig7D}
\gamma^{2}(\omega) &=& \frac{(2/\pi)^{1/2}\Gamma(3/4)\sin(\pi/8)}{\vert\omega\vert^{3/4}}. \\ \nonumber
\textrm{Where} \quad
\Gamma(3/4) &=& \Big(\frac{\pi}{2}\Big)^{1/4}\bigg[ \sum_{k = -\infty}^{\infty} e^{-2\pi k^{2}} \theta_{4}(ik\pi,e^{-\pi}) \bigg]^{-1/2}, 
\end{eqnarray}

and $\theta_{4}$ is the Jacobi elliptic function \cite{Mezo}.
\subsection{Case 4: GHM model (Golla, Hughes, and McTavish)}

Similar to the Biot model, the Golla, Hughes, and McTavish model \cite{Hughes1,Hughes2} has a material modulus function, 
or the memory kernel function comprising of a set of mini-oscillators each indexed by $k = 1,2,...,n$. It has has the 
following set of real constants $\{\alpha_{k},b_{1k},b_{2k}\}$, where $\alpha_{k} > 0$, and $b_{2k} > b_{1k} > 0$. If we choose  

\begin{eqnarray}\label{Zwanzig8C}
w(\omega) &=& 2\omega^{2}, \quad a_{k} = \frac{\alpha_{k}}{b_{2k} - b_{1k}}, \\ \nonumber
\textrm{and} \quad 
\gamma^{2}(\omega) &=& \sqrt{\frac{2}{\pi}}\sum_{k=1}^{n} a_{k}\bigg{\{} \frac{1}{b^{2}_{1k} + \omega^{2}} - 
\frac{1}{b^{2}_{2k} + \omega^{2}} \bigg{\}}, \\ \nonumber
\end{eqnarray}

then the memory kernel function in the time domain will be given by   

\begin{equation}\label{Zwanzig8D}
G(t) = \sum_{k=1}^{n} \alpha_{k}\frac{(b_{2k}e^{-b_{1k}t} - b_{1k}e^{-b_{2k}t})}{b_{2k} - b_{1k}}.   
\end{equation}

The Zwanzig model offers a simple interpretation for the GHM case. The environment excites the system over a range of frequencies 
given by the sets $\{2\pi b_{1,2k}\}$ and weighted by $\{a_{k}b_{2,1k}\}$ respectively. Depending upon the lifetime of the excitations 
$b^{-1}_{1,2k}$, the system will acquire a memory due to the environment given by the kernel function in Eq.\eqref{Zwanzig8D}. 

\subsection{Case 5: Gaussian model (Adhikari and Woodhouse)}

The simple Gaussian case by Adhikari and Woodhouse \cite{Adhikari1} can be obtained by choosing  

\begin{equation}\label{Zwanzig8E}
w(\omega) = 2\omega^{2}, \quad \textrm{and} 
\quad \gamma^{2}(\omega) = \frac{\exp(-\omega^{2}/4\mu)}{\sqrt{2\mu}}. 	 	
\end{equation}

With these choices, the memory kernel function in Eq.\eqref{Zwanzig9} will give us 

\begin{equation}\label{Zwanzig8F}
G(t) = \frac{1}{\sqrt{2\mu}} \int_{-\infty}^{\infty}d\omega e^{-\omega^{2}/4\mu} e^{i\omega t} = \exp(-\mu^{2}t^{2}).	 	
\end{equation}

\section{Summary}
The harmonic oscillator bath model for the environment is a well-established theory, developed using a first-principles-based approach. 
The present study demonstrates that the classical viscous damping model and various non-viscous damping models proposed for different 
engineering applications can be subsumed within this model. For specific choices of the memory kernel function, exact expressions for 
the damping coefficient of a wide range of damping models can be derived. The study traces seemingly ad-hoc phenomenological models of 
damping to theories grounded in the principles of statistical mechanics, thereby providing deeper physical insights into a broad class 
of engineering models of non-viscous damping. 

Designing an environment through carefully engineered excitations presents a significant challenge in material design. We believe that 
Zwanzig’s approach to modeling such an environment represents an important theoretical advancement in this field. We examined the dynamic 
aspects of damping, specifically where the memory kernel function depends solely on time or frequency. However, this formulation provides 
a general framework that can be expanded to include spatial considerations as well. To effectively model environments in the spacetime 
continuum, we need to define field variables that are continuous functions of both space and time. This area is currently the focus of 
ongoing research, and we will share our findings at a later date.
\section{Appendix}

In this Appendix section, we will review some basic results in classical Hamiltonian mechanics and statistical mechanics
which has been used in this paper, and derive the fluctuation dissipation theorem.

\subsection{Hamilton's equation of motion and the Poisson bracket}

For a system with position variables $X$, its canonically conjugate momentum variables $P$, and Hamiltonian $H(X,P)$, 
the Hamilton's equations of motion are:

\begin{equation}\label{App1}
\frac{dX}{dt} = \frac{\partial H}{\partial P}, \quad  
\frac{dP}{dt} = -\frac{\partial H}{\partial X}, \quad  
\end{equation}

For any canonical variable $A = A(X,P)$, the equation of motion:

\begin{eqnarray}\label{App2}
\frac{dA}{dt} &=& \frac{\partial A}{\partial X}\frac{\partial X}{\partial t} +  
\frac{\partial A}{\partial P} \frac{\partial P}{\partial t}   \\ \nonumber 
\frac{dA}{dt} &=& \frac{\partial A}{\partial X}\frac{\partial H}{\partial P} -  
\frac{\partial A}{\partial P} \frac{\partial H}{\partial X} = 
\{A,H\}_{PB}  
\end{eqnarray}

The second equation follows from the Eq.\eqref{App1}, and

\begin{equation}\label{App3}
\{A,H\}_{PB} = 
\frac{\partial A}{\partial X}\frac{\partial H}{\partial P} -  
\frac{\partial A}{\partial P}\frac{\partial H}{\partial X}  
\end{equation}

is known as the Poisson bracket. Using this definition, the equations of motion in Eqs.\eqref{Zwanzig0}-\eqref{Zwanzig5A} can be derived 
accordingly.
\subsection{The equilibrium distribution of the harmonic oscillators and the statistical averages}

Let us consider the following integral:

\begin{eqnarray}\label{App4}
Z_{b} &=& \prod_{i} \int_{-\infty}^{+\infty} \int_{-\infty}^{+\infty} dX_{i} dP_{i} \exp(-H_{b}/k_{B}T), \\ 
\textrm{where,} \quad
H_{b} &=& \sum_{j=1}^{N}\bigg[\frac{P_{j}^{2}}{2} + \frac{\omega^{2}_{j}}{2}\bigg(X_{j} - \frac{\gamma_{j}}{\omega^{2}_{j}}x\bigg)^{2} \bigg]   
\quad (\textrm{Eq}.\eqref{Zwanzig1}).
\end{eqnarray}

The oscillators are identically and independently distributed, where the distribution function for each oscillator is a Gaussian in ($X_{i},P_{i}$), 
with means ($\gamma_{j}\omega^{-2}_{j}x,0$), and variance $k_{B}T$. Therefore, the normalization constant $Z_{b} = (\pi k_{B}T)^{N}$. 
Using the definition in Eq.\eqref{App6} and the probability density functions in Eq.\eqref{App4}, we obtain the following standard results in 
Eq.\eqref{Zwanzig7}:

\begin{eqnarray}\label{App8}
&&\langle P_{i} \rangle = 0, \quad \langle P_{i}P_{j}\rangle = \delta_{ij}k_{B}T, \\ \nonumber
&&\Big\langle (X_{i}-\gamma_{i}\omega^{-2}_{i}x)(X_{j}-\gamma_{j}\omega^{-2}_{j}x)\Big\rangle = \delta_{ij}k_{B}T/\omega^{2}_{j}, \\ \nonumber
\textrm{and,} \quad
&&\Big\langle P_{i}(X_{j}-\gamma_{j}\omega^{-2}_{j}x)\Big\rangle = \langle P_{i}\rangle \Big\langle (X_{j}-\gamma_{j}\omega^{-2}_{j}x)\Big\rangle = 0
\end{eqnarray}
\subsection{The fluctuation dissipation theorem}

The first equation in \eqref{Zwanzig5B} gives the random force due to the environment.
From the results obtained in the previous section, we see that $\langle F(t) \rangle = 0$.
Where, we have used the fact that $\langle P_{i} \rangle = 0$ and $\langle X_{i} \rangle = \gamma_{i}\omega^{-2}_{i}x$.
Using the results in Eq.\eqref{App8}, we will now derive the fluctuation-dissipation theorem in Eq.\eqref{Zwanzig8}.

\begin{eqnarray}\label{App7}
\langle F(t)F(t')\rangle &=& \sum_{i,j=1}^{N}
\gamma_{i}\gamma_{j}\bigg[\underbrace{\langle P_{i}(0)P_{j}(0)\rangle}_{\delta_{ij}k_{B}T}
\frac{\sin\omega_{i}t \sin\omega_{j}t'}{\omega_{i}\omega_{j}} \\ \nonumber
&+& \underbrace{\Big\langle (X_{i}(0) - \gamma_{i}\omega^{-2}_{i}x(0))(X_{j}(0) - \gamma_{j}\omega^{-2}_{j}x(0))\Big\rangle}_{\delta_{ij}k_{B}T/\omega^{2}_{j}}
\cos\omega_{i}t \cos\omega_{j}t' \\ \nonumber
&-& 2\underbrace{\Big\langle P_{i}(0)(X_{j}(0) - \gamma_{j}\omega^{-2}_{j}x(0))\Big\rangle}_{0} \sin\omega_{i}t \cos\omega_{j}t' \bigg].
\end{eqnarray}

\begin{equation}\label{App9}
\textrm{Or,} \quad 
\langle F(t)F(t')\rangle = k_{B}T \sum_{j=1}^{N}\gamma^{2}_{j}\frac{\cos\omega_{j}(t-t')}{\omega^{2}_{j}} = k_{B}T G(t-t').
\end{equation}

Where we have used the trigonometric identity $\cos(C-D) = \cos C \cos D + \sin C \sin D$. When $N \rightarrow \infty$, the frequency 
$\omega_{j} \rightarrow \omega$, and $\gamma_{j} \rightarrow \gamma(\omega)$. If $w(\omega)$ be the density of oscillators, then in the 
interval between $\omega$ and $\omega+d\omega$, we will have $w(\omega)d\omega$ oscillators. In this limit, $G(t)$ in Eq.\eqref{Zwanzig5B} 
will be

\begin{eqnarray}\label{App10}
\sum_{j=1}^{N}\gamma^{2}_{j}\frac{\cos\omega_{j}(t-t')}{\omega^{2}_{j}} &=& 
\int d\omega w(\omega)\gamma^{2}(\omega)\frac{\cos\omega(t-t')}{\omega^{2}}. \\ \nonumber 
\textrm{Or,} \quad 
G(t-t') &=& \int_{-\infty}^{\infty}d\omega w(\omega)\frac{\gamma^{2}(\omega)}{2\omega^{2}} e^{i\omega t}, \quad \textrm{in general}.             
\end{eqnarray}

The equal time correlation or the fluctuations $\langle F^{2}(t)\rangle = G(0)k_{B}T$, where $G(0)$ is some constant involving a 
frequency dependent cut-off. 

\noindent
{\bf CRediT authorship contribution statement}

SKG: Calculations, developing theory, writing, review and editing - original draft.

IM: Calculations, developing theory, writing, review and editing - original draft.

\noindent
{\bf Funding}: This research did not receive any specific grant from funding agencies in the public, commercial, or not-for-profit sectors.

\noindent
{\bf Data Availability Statement}: No Data associated in the manuscript.

\noindent
{\bf Conflict of Interest}: The authors declare that they have no known competing financial interests or personal relationships that could have 
appeared to influence the work reported in this paper.


\end{document}